\documentclass[twocolumn,showpacs,preprintnumbers,amsmath,amssymb]{revtex4}

\usepackage{graphicx}
\usepackage{dcolumn}
\usepackage{bm}
\usepackage{mathrsfs}
\usepackage{amssymb}
\usepackage{amsmath}
\usepackage{amsfonts}
\usepackage{amsfonts,enumerate}
\usepackage{pifont}
\usepackage{enumerate}
\usepackage{graphics}
\usepackage{verbatim}
\usepackage{amsthm}

\begin{document}


\title{Geometrical perspective on quantum states and quantum computation}

\author{Zeqian Chen}
\affiliation{%
State Key Laboratory of Resonances and Atomic and Molecular Physics, Wuhan Institute of Physics and Mathematics,
Chinese Academy of Sciences, 30 West District, Xiao-Hong-Shan, Wuhan 430071, China}%


\begin{abstract}
We interpret quantum computing as a geometric evolution process by reformulating finite quantum systems via Connes' noncommutative geometry. In this formulation, quantum states are represented as noncommutative connections, while gauge transformations on the connections play a role of unitary quantum operations. Thereby, a geometrical model for quantum computation is presented, which is equivalent to the quantum circuit model. This result shows a geometric way of realizing quantum computing and as such, provides an alternative proposal of building a quantum computer.
\end{abstract}

\maketitle


Quantum computation has the advantage of solving efficiently some problems that are considered intractable by using conventional classical computation \cite{NC2000}. In this context, there are two remarkable algorithms found: Shor's factoring algorithm \cite{Shor1994} and Grove's search algorithm \cite{Gro1997}. But it remains a challenge to find efficient quantum circuits that can perform these complicated tasks in practice, due to quantum decoherence. A crucial step in the theory of quantum computer has been the discovery of error-correcting quantum codes \cite{Shor1995} and fault-tolerant quantum computation \cite{Preskill1997, Shor1996}, which established a threshold theorem that proves that quantum decoherence can be corrected as long as the decoherence is sufficiently weak. To tackle this barrier, a revolutionary strategy, topological quantum computation (see \cite{NSSFD2008} and references therein), is to make the system immune to the usual sources of quantum decoherence, by involving the globally robust topological nature of the computation. Recently, substantial progress in this field has been made on both theoretical and experimental fronts \cite{SL2013}.

In this paper, we provide an alternative approach to quantum computation from a geometrical view of point. To this end, we need to reformulate quantum mechanics via Connes' noncommutative geometry \cite{Connes1994}. In this formulation, quantum states are represented as noncommutative connections, while gauge transformations on the connections play a role of unitary quantum operations. In this way, we present a geometrical model for quantum computation, which is equivalent to the quantum circuit model. In this computational model, information is encoded in gauge states instead of quantum states and implementing on gauge states is played by gauge transformations. Therefore, our scheme shows a geometric way of realizing quantum computing and as such, provides an alternative proposal of building a quantum computer.


Let $\mathbb{H}$ be a $N$ dimensional Hilbert space associated with a finite quantum system. Let $\mathcal{A}$ be the algebra of all (bounded) linear operators on $\mathbb{H},$ and let $\mathcal{U} (\mathcal{A}) = \{ u \in \mathcal{A}: u u^* = u^* u = I\}$ with $I$ being the unit operator on $\mathbb{H}.$ Given a selfadjoint operator $D$ on $\mathbb{H},$ $(\mathcal{A}, \mathbb{H}, D)$ is a spectral triple in the sense of noncommutative geometry \cite{Connes1994, Landi1997}. A (noncommutative) connection on $(\mathcal{A}, \mathbb{H}, D)$ is defined to be a selfadjoint operator $V$ on $\mathbb{H}$ of the form that follows
\begin{equation}\label{eq:Connection}
V = \sum_j a_j [D, b_j]
\end{equation}
where $a_j, b_j \in \mathcal{A}$ and $[a, b] = a b - b a.$ A gauge transform on a connection $V$ under $u \in \mathcal{U} (\mathcal{A})$ is defined as
\begin{equation}\label{eq:Gaugetransf}
V \longmapsto G_u (V) = u V u^* + u [D, u^*].
\end{equation}
For avoiding triviality, we always assume that $D \neq 0$ or $I$ in what follows.

For any (pure) quantum state $| \psi \rangle \langle \psi|$ with $\psi$ being a unit vector in $\mathbb{H},$ we have
$$
| \psi \rangle \langle \psi| = | \psi \rangle \langle \varphi | \mathrm{i} [D, b] | \varphi \rangle \langle \psi|
$$
where $\mathrm{i} = \sqrt{-1}$ and, $b$ is a selfjoint operator on $\mathbb{H}$ such that $\mathrm{i} [D, b]$ has eigenvalue $1$ at $| \varphi \rangle.$ Such a selfjoint operator $b$ always exists because $D \neq 0$ or $I.$ In this case,
\begin{equation}\label{eq:statebyconnection}
| \psi \rangle \langle \psi| = \mathrm{i} a^* [D, b a] - \mathrm{i} a^* b [D, a]
\end{equation}
with $a =  | \varphi \rangle \langle \psi|.$ Thus, every quantum state $| \psi \rangle \langle \psi|$ can be represented as a connection, denoted by $V_{\psi},$ i.e., \begin{equation}\label{eq:purestateConnection}
V_{\psi} = \mathrm{i} a^* [D, b a] - \mathrm{i} a^* b [D, a].
\end{equation}

Let $\mathcal{G}_D (\mathbb{H})$ be the set of all connections $V$ which can be written as $V = V_\psi + u D u^* -D$ with $\psi$ being a unit vector in $\mathbb{H}$ and $u \in \mathcal{U}(\mathcal{A}).$ An element in $\mathcal{G}_D (\mathbb{H})$ is said to be a {\it gauge state} on $(\mathcal{A}, \mathbb{H}, D).$ Any quantum state is necessarily a gauge state, but a gauge state need not to be a quantum state. However, any gauge state $V$ can be obtained from a quantum state by performing a gauge transform. Indeed, if $V = V_\psi + u D u^* -D$ then $V = G_u (V_{u^* \psi}).$ Moreover, for any gauge state $V$ on $(\mathcal{A}, \mathbb{H}, D)$ we have (see \cite{A1})
\begin{itemize}

\item for any $u \in \mathcal{U}(\mathcal{A}),$ $G_u (V)$ is again a gauge state;

\item $G_{u v} (V) = G_u (G_v(V))$ for all $u, v \in \mathcal{U}(\mathcal{A}).$

\end{itemize}
Therefore, a gauge transform preserves gauge states.

Let $V$ be a gauge state which is prepared from a quantum state $| \psi \rangle \langle \psi|$ by operating a gauge transform $G_u,$ i.e., $V = G_u (V_\psi).$ For any event $E,$ the probability of $E$ occurring on $V$ is
\begin{equation}\label{eq:Probgauge}
\langle E \rangle_V = \langle \psi | u^* E u | \psi \rangle.
\end{equation}
Note that a gauge state may be prepared in several ways. Hence, the probability of a event $E$ occurring on a gauge state $V$ depends on the quantum state from which $V$ is prepared.

Let $H$ be a selfadjoint operator on $\mathbb{H}.$ Assuming $u_t = e^{\mathrm{i} t H}$ for $t \in \mathbb{R},$ we have that the gauge transforms $V_t = G_t (V)$ on a fixed gauge state $V$ under $u_t$ form a group (see \cite{A1}), that is,
\begin{equation}\label{eq:dynamicalgroup}
G_{t +s} (V) = G_t (G_s (V)).
\end{equation}
This yields a dynamical equation governed by the Hamiltonian $H$ for gauge states on  $(\mathcal{A}, \mathbb{H}, D)$ as follows \cite{A2}
\begin{equation}\label{eq:dynamicalequation}
\mathrm{i} \frac{d V_t}{ d t} = [V_t, H] + [D, H]
\end{equation}
with $V_0 =V.$ In particular, for a unit vector $\psi$ we have
\begin{equation}\label{eq:grounddynamic}
V_t = G_t (V_\psi) = V_{ u_t \psi} + u_t D u^*_t - D.
\end{equation}

We now turn to product of two spectral triples. Suppose $(\mathcal{A}_i, \mathbb{H}_i, D_i),$ $i=1,2,$ are two spectral triple associated with finite quantum systems. Put
\begin{equation}\label{eq:Dcompo}
D = D_1 \otimes I_2 + I_1 \otimes D_2
\end{equation}
with $I_i$ being the unit operator on $\mathbb{H}_i$ ($i =1,2$). Then $D$ is a selfjoint operator on $\mathbb{H}_1 \otimes \mathbb{H}_2.$ The spectral triple $(\mathcal{A}_1 \otimes \mathcal{A}_2, \mathbb{H}_1 \otimes \mathbb{H}_2, D)$ is called the product of two spectral triples $(\mathcal{A}_i, \mathbb{H}_i, D_i),$ $i =1,2.$

Now we illustrate our scheme by using a qubit. Let $\mathbb{H} = \mathbb{C}^2$ and
\begin{equation}\label{eq:Dqbit}
\sigma_x = \begin{bmatrix}
0 & 1 \\
1 & 0
\end{bmatrix},
\sigma_y = \begin{bmatrix}
0 & - \mathrm{i} \\
\mathrm{i} & 0
\end{bmatrix},
\sigma_z = \begin{bmatrix}
1 & 0 \\
0 & -1
\end{bmatrix}.
\end{equation}
Then $(\mathbb{M}_2, \mathbb{C}^2, D)$ is a spectral triple with $D = \sigma_x,$ where $\mathbb{M}_2$ is the set of all $2 \times 2$ complex matrices. For $|0\rangle = \begin{bmatrix}
1 \\
0
\end{bmatrix},$ we have
\begin{equation*}
V_{|0\rangle} = \begin{bmatrix}
1 & 0 \\
0 & 0
\end{bmatrix}, \quad
G_{\sigma_x} (V_{|0\rangle}) = \begin{bmatrix}
0 & 0\\
0 & 1
\end{bmatrix},
\end{equation*}
and
\begin{equation*}
G_{\sigma_y} (V_{|0\rangle}) = \begin{bmatrix}
0 & -2\\
-2 & 1
\end{bmatrix},\quad
G_{\sigma_z} (V_{|0\rangle}) = \begin{bmatrix}
1 & -2\\
-2 & 0
\end{bmatrix}.
\end{equation*}
For $|1\rangle = \begin{bmatrix}
0 \\
1
\end{bmatrix},$ we have
\begin{equation*}
V_{|1\rangle} = \begin{bmatrix}
0 & 0 \\
0 & 1
\end{bmatrix}
\quad \text{and}\quad
G_{\sigma_y} (V_{|1\rangle}) = \begin{bmatrix}
1 & -2\\
-2 & 0
\end{bmatrix}.
\end{equation*}
Hence $G_{\sigma_y} (V_{|1\rangle}) = G_{\sigma_z} (V_{|0\rangle})$ and so, the gauge state
\begin{equation*}
V= \begin{bmatrix}
1 & -2\\
-2 & 0
\end{bmatrix}
\end{equation*}
can be prepared in two different ways.


We are now ready to interpret quantum computation from a geometrical view of point. But let us take a step backward and discuss the standard quantum circuit model for computation \cite{Gudder2003}. Let $\mathbb{H} = (\mathbb{C}^2)^{\otimes^n},$ the tensor product of $n$ copies of $\mathbb{C}^2.$ A quantum circuit model on $n$ qubits consists of
\begin{itemize}

\item a initial state $| \psi \rangle,$ represented by a unit vector $\psi \in \mathbb{H};$

\item a quantum circuit $\Gamma = U_N U_{N-1} \cdots U_1,$ where quantum ``gates" $U_k$ $1 \leq k \leq N,$ are unitary transformations on either $\mathbb{C}^2_i$ or $\mathbb{C}^2_i \otimes \mathbb{C}^2_j,$ $1 \leq i, j \leq n,$ the identity on all remaining factors;

\item reading the output of the circuit $\Gamma | \psi \rangle$ by measuring the first qubit; the probability of observing $|1 \rangle$ is $P(\Gamma) = \langle \psi | \Gamma^* \Pi_1 \Gamma | \psi \rangle,$ where $\Pi_1 = | 1\rangle \langle 1| \otimes I \cdots \otimes I$ is the projection to $| 1\rangle$ in the first qubit.

\end{itemize}

Let $\mathcal{A} = \mathbb{M}_{2^n}.$ Put
\begin{equation*}
D = \sum_{i=1}^n \underbrace{I \otimes \cdots \otimes I}_{i-1} \otimes \sigma_x \otimes I \cdots \otimes I
\end{equation*}
where $I$ is the identity on $\mathbb{C}^2.$ A computational model based on the spectral triple $(\mathcal{A}, \mathbb{H}, D)$ is as follows:
\begin{itemize}

\item Initialization of a gauge state $V_\psi$ in the spectral triple $(\mathcal{A}, \mathbb{H}, D),$ where $\psi$ is a unit vector in $\mathbb{H};$

\item Gauge implementation of the computational program
\begin{equation*}
G(\Gamma) = G_{U_N} G_{U_{N-1}} \cdots G_{U_1}
\end{equation*}
where ``gates" $G_{U_k},$ $1 \leq k \leq N,$ are gauge transformations induced by $U_k;$

\item Application of the projection operator $\Pi_1$ for reading the output of the computation $G(\Gamma) (V_\psi);$ the probability of observing $|1 \rangle$ is $P(G_\Gamma) = \langle \psi | \Gamma^* \Pi_1 \Gamma | \psi \rangle$ because $G(\Gamma) (V_\psi) = G_{\Gamma} (V_\psi)$ (see \cite{A1}), i.e., $G(\Gamma) (V_\psi) = \Gamma | \psi \rangle \langle \psi | \Gamma^* + \Gamma D \Gamma^* -D.$

\end{itemize}
Thus, we obtain a geometrical model on $n$ qubits for quantum computation, which is evidently equivalent to the quantum circuit model as described above. Due to the essential role of gauge transformations played in this computational model, we call this scheme {\it gauge quantum computation}.

As illustration, we give the Deutsch-Jozsa algorithm \cite{DJ1992} in gauge quantum computation. Let $f: \{0, 1\}^n \mapsto \{0, 1\}$ be a function that takes an $n$-bit into a bit. We call $f$ balanced if $f(x) = 1$ for exactly half of all possible $x$ and $f(x) =0$ for the other half. Given a function $f$ that is either constant or balanced, we want to find out which it is with certainty. More precisely, we select one $x \in \{0, 1\}^n$ and calculate $f(x)$ with the result being either $0$ or $1.$ What is the fewest number of queries that we can make to determine whether or not $f$ is constant? In the classical case, at worst we will need to calculate $f$ $2^{n-1}+1$ times, because we may first obtain $2^{n-1}$ zeros and will need one more query to decide. However, in the setting of quantum computation we could achieve the goal in just one query using the Deutsch-Jozsa algorithm. In the sequel, we give a realization of the Deutsch-Jozsa algorithm in gauge quantum computation.

Let $\mathbb{H} = (\mathbb{C}^2 )^{\otimes (n+1)}$ and $\mathcal{A} = \mathbb{M}_{2^{n+1}}.$ Given a selfadjoint operator $D$ on $\mathbb{H}$ that is not $0$ or $I,$ we get the desired spectral triple $(\mathcal{A}, \mathbb{H}, D).$ For a given $f,$ we define the associated operator $U_f $ on $\mathbb{H}$ as $U_f | x, y \rangle = | x , y \oplus f(x) \rangle$ for $ x \in \{0,1\}^n$ and $y \in \{0,1 \}.$ Recall that the Hadamard operator $H$ on $\mathbb{C}^2$ is
\begin{equation*}
H = \frac{1}{\sqrt{2}} \sum_{x, y \in \{0, 1\}} (-1)^{x \cdot y} | x \rangle \langle y|
\end{equation*}
where $x \cdot y$ signifies ordinary multiplication. The following is the Deutsch-Jozsa algorithm in the setting of gauge quantum computation:
\begin{itemize}

\item Initialization of a gauge state $V_\psi$ with $\psi = | 0 \rangle^{\otimes n}\otimes | 1 \rangle;$

\item Gauge implementation of the computational program $G(\Gamma) = G_{H^{\otimes n} \otimes I} G_{U_f} G_{H^{\otimes (n+1)}};$

\item Application of the projection operator $\Pi_{|0\rangle^{\otimes n}}$ for reading the output of the computation $G(\Gamma) (V_\psi),$ where $\Pi_{|0\rangle^{\otimes n}}$ is the projection to $| 0 \rangle^{\otimes n}$ in the first $n$ qubits.

\end{itemize}
The final gauge state is $V = V_{\Gamma \psi} + \Gamma D \Gamma^* -D$ with $\Gamma = (H^{\otimes n} \otimes I) U_f H^{\otimes (n+1)},$ where
\begin{equation*}
\Gamma \psi = \sum_{x, y \in \{0, 1\}^n} \frac{(-1)^{x \cdot y + f(x)}}{2^n} | y \rangle \otimes \frac{|0\rangle - |1\rangle}{\sqrt{2}}.
\end{equation*}
Since the amplitude for the state $|0 \rangle^{\otimes n}$ in the first $n$ qubits is $\sum_x (-1)^{f(x)}/ 2^n,$ the probability of observing $0$ is $1$ if $f$ is constant, or $0$ if $f$ is balanced. Thus we have two possibilities of obtaining the outcome zero or the outcome nonzero. In the first case, $f$ is certainly constant and in the second case $f$ must be balanced. Therefore, we only need to perform three times gauge transforms for determining whether or not $f$ is constant.

In conclusion, we present a geometrical description of quantum computation via noncommutative geometry. In this geometrical model, information is encoded in gauge states and computational operation is implemented by gauge transforms instead of unitary transforms. In principle, gauge transforms are easier to perform than unitary quantum operation \cite{Madore1999}. Therefore, gauge quantum computation should be more accessible than the usual quantum circuit computation and as such, this provides an alternative proposal of building a quantum computer.

This work was supported in part by the NSFC under Grant No. 11171338 and National Basic Research Program of China under Grant No. 2012CB922102.

\bibliography{apssamp}

\begin{thebibliography}{**}
\bibitem{A1} Since $G_u (V) = u V u^* + u D u^* - D,$ it follows that
\begin{equation*}\begin{split}
G_u (G_v (V)) & = u (v V v^* + v D v^* -D)u^* + u D u^* - D\\
& = u v V (u v)^* + uv D (u v)^* - D = G_{u v} (V).
\end{split}\end{equation*}
\bibitem{A2} Since
\begin{equation*}\begin{split}
\frac{d G_t (V)}{ d t} & = \frac{ d (u_t V u^*_t)}{d t} + \frac{ d (u_t D u^*_t)}{d t}\\
& = \mathrm{i} [H, u_t V u^*_t] + \mathrm{i} [H, u_t D u^*_t]= \mathrm{i} [H, V_t +D],
\end{split}\end{equation*}
it follows \eqref{eq:dynamicalequation}.
\end{thebibliography}

\end{document}